\documentclass[10pt,conference]{IEEEtran}
\IEEEoverridecommandlockouts

\usepackage{cite}
\usepackage{amsmath,amssymb,amsfonts}
\usepackage{graphicx}
\usepackage{booktabs}
\usepackage{multirow}
\usepackage{textcomp}
\usepackage{xcolor}
\usepackage{xspace}
\usepackage{listings}
\usepackage{caption}
\usepackage{dblfloatfix}
\usepackage[override]{cmtt}

\newenvironment{fixedlisting}[2]%
  {\par\addvspace{\bigskipamount}\noindent
   \def\fixedlistingcap{#1}\def\fixedlistinglab{#2}%
   \begin{minipage}{\columnwidth}}%
  {\captionof{lstlisting}{\fixedlistingcap}\label{\fixedlistinglab}%
   \end{minipage}\par\addvspace{\bigskipamount}}

\definecolor{lstkw}{rgb}{0.13,0.13,0.55}      
\definecolor{lstcomment}{rgb}{0.28,0.48,0.28} 
\definecolor{lststring}{rgb}{0.55,0.20,0.10}  
\definecolor{lstnum}{rgb}{0.50,0.50,0.50}     
\definecolor{lstframe}{rgb}{0.75,0.75,0.75}   
\definecolor{lstbg}{rgb}{0.985,0.985,0.985}   
\definecolor{diffadd}{rgb}{0.00,0.45,0.00}    
\definecolor{diffdel}{rgb}{0.70,0.10,0.10}    

\lstdefinestyle{javaStyle}{language=Java}
\lstdefinestyle{scalaStyle}{language=Scala}

\lstdefinelanguage{json}{
    morestring=[b]",
    showstringspaces=false,
}
\lstdefinestyle{jsonStyle}{language=json}

\lstdefinelanguage{JavaDiff}{
  language=Java,
  moredelim=[l][\color{diffdel}]{-},
  moredelim=[l][\color{diffadd}]{+},
}
\lstdefinestyle{diffStyle}{language=JavaDiff}

\lstdefinelanguage{ScalaDiff}{
  language=Scala,
  moredelim=[l][\color{diffdel}]{-},
  moredelim=[l][\color{diffadd}]{+},
}
\lstdefinestyle{scalaDiffStyle}{language=ScalaDiff}

\begin{document}

\title{Agentic Root Cause Reasoning and Localization of Unsoundness in Static Call Graph Analysis}


\author{\IEEEauthorblockN{Mohammad Rafieian}
\IEEEauthorblockA{
\textit{The University of Texas at Dallas}\\
Richardson, Texas, USA \\
mohammad.rafieian@utdallas.edu}
\and
\IEEEauthorblockN{Michael Hicks}
\IEEEauthorblockA{\textit{University of Pennsylvania and Amazon} \\
Philadelphia, Pennsylvania, USA \\
mwh@cis.upenn.edu}
\and
\IEEEauthorblockN{Shiyi Wei}
\IEEEauthorblockA{
\textit{The University of Texas at Dallas}\\
Richardson, Texas, USA \\
swei@utdallas.edu}
}


\newcommand{\moh}[1]{\textcolor{blue}{#1}}
\newcommand{\sw}[1]{\textcolor{red}{#1}}
\newcommand{\mwh}[1]{\textcolor{green}{#1}}

\newcommand{\tool}{CG\textit{Rizz}\xspace}
\newcommand{\training}{training\xspace}
\newcommand{\Training}{Training\xspace}

\maketitle

\begin{abstract}

Static call graph construction tools cannot soundly model every dynamic language feature, yet they rarely reveal when, where, or why their results are unsound. Users cannot judge whether a call graph is complete for their program, and developers lack the precise root cause and code location needed to tell an implementation bug from an intended soundness/precision tradeoff. 
We present \tool, an agentic framework addressing both needs in two phases. In a \emph{\training} phase, \tool compares a benchmark of programs' static and dynamic call graphs to find \emph{boundary edges}---calls observed at runtime but missing from the static graph---and, for each, an LLM agent suggests a root cause: the program pattern the analysis failed to model and the responsible location in the tool. Because such a suggestion may be wrong, \tool accepts it only when three independent pieces of evidence agree: runtime instrumentation that detects the pattern, a test case that reproduces the missed edge, and a warning inserted at the implicated tool location. The validated reports, test cases, and code locations serve developers; the warnings accumulate into a warning-instrumented build that, in a \emph{deployment} phase, flags unsoundness on new programs for users, with no dynamic analysis required.
We evaluate \tool with three state-of-the-art tools, WALA, OPAL, and Soot, on three complementary benchmark suites. Of the 1459 boundary edges \tool encounters, it reasons about 10.3\% directly, all confirmed by a manual audit, and resolves a further 84.1\% by grouping them with these validated root causes. Tool maintainers confirmed several reported limitations as genuine, and some have been fixed. The generated warnings generalize, flagging 74\%--92\% of the unsoundness in held-out programs. Finally, guided by \tool’s reports, a coding agent refactored programs to remove 64\% of the targeted root-cause patterns while preserving functionality, making them more amenable to static analysis.
\end{abstract}

\begin{IEEEkeywords}
call graphs, root cause analysis, agent reasoning.
\end{IEEEkeywords}

\vspace{-3pt}
\section{Introduction}
\label{sec:intro}

A \emph{static call graph} models the potential calling relationships between methods in a program. Such a graph has traditionally served as the foundation of a wide range of program analysis applications, including vulnerability detection, program comprehension, optimization, and software maintenance~\cite{v1, v2, v3}. Today, call graphs also provide useful code context to large language models (LLMs), enabling them to reason about software behavior, answer developer queries, and perform complex software engineering tasks~\cite{cgllm1, cgllm2}.

Ideally, static call graph construction tools are \emph{sound}, meaning that they capture all possible calling relationships that may occur during any execution of the target program. In practice, achieving soundness for all possible programs is rarely possible~\cite{Sui_2020}. This limitation often stems from dynamic language features such as reflection, native code, and other behaviors that are difficult---or in some cases impossible---to analyze statically~\cite{Sui_2020, Reif_2018, Helm_2024, Samhi_2024}.

Tools that consume unsound call graphs may produce wrong answers. For example, a taint analysis tool~\cite{taint_analysis} that does not see a call to a particular method may miss an illicit flow (leading to a false negative), or an exonerating input validation (leading to a false positive). Especially in the former case, users (including LLM coding agents) may be overconfident that their program is free of certain classes of vulnerabilities.

Unsound analysis tools leave two groups of people missing useful information. A \emph{user}, who runs a tool on their own program, has no way to tell \emph{whether, where, and why} the resulting call graph is unsound for that program: documentation offers only aggregate statistics, and the tool itself issues no warning when it gives up on a construct. A \emph{developer}, who maintains the tool, lacks the precise root cause and code location needed to tell a genuine implementation bug from an intended soundness/precision tradeoff---and, when it is a bug, to reproduce and fix it.

We developed \tool to address these unmet needs. It operates in two phases. In a \emph{\training} phase, \tool compares a benchmark of programs' static and dynamic call graphs to find \emph{boundary edges}---calls observed during execution but absent from the static graph---and, for each, an LLM agent \emph{suggests} a root cause: the program pattern the analysis failed to model and the responsible location in the tool's code. Because the suggestion can be wrong, \tool accepts it only when three independent pieces of evidence agree: runtime instrumentation that catches the pattern as the program executes, a test case that reproduces the missed edge, and a warning inserted at the implicated tool location. The resulting report, reproducing test case, and code location are what a developer needs to confirm, fix, or knowingly accept a limitation. The inserted warnings accumulate into a single warning-instrumented build; in a \emph{deployment} phase, this build is run by a user on a new program and warns where and why that program's call graph may be missing edges, giving a user a view of whether to trust the analysis or adjust the program.

 
We evaluate \tool on three widely used static call graph construction frameworks---WALA, OPAL, and Soot---using three complementary benchmarks, asking whether \tool produces trustworthy artifacts and whether those artifacts serve the two stakeholders. First, of the 1459 boundary edges the \training phase encounters, an agent reasons about and validates 151 directly, and a manual audit confirms the reported root cause in every one of them; grouping resolves a further 1227 by matching them to an already-validated root cause, and re-analyzing a stratified sample shows that 86\% of grouped edges share their group's root cause. Second, the resulting artifacts are actionable for tool developers: of the reports we shared with the maintainers of each tool, several were confirmed as genuine limitations, and some have been fixed. Third, the warnings learned during \training generalize to previously unseen programs: in a cross-validation study they surface 74\%--92\% of the unsoundness in held-out programs, and running the warning-instrumented tools directly on eight new programs flags 75\%--83\% of their root causes, giving a user a soundness signal for their own code with no dynamic analysis required. Finally, a user can act on these reports to reduce unsoundness: guided by \tool’s reports, a coding agent refactored programs to remove root-cause patterns while preserving their functionality, resolving 30 of 47 boundaries (64\%). 




Overall, this paper offers the following contributions:
\begin{itemize}
\item We present \tool, the first agentic framework that reasons about, validates, and localizes the root causes of unsoundness in static call graph construction. Prior work characterizes a tool's unsoundness only in aggregate---by comparing static and dynamic call graphs~\cite{chakraborty_2022, Sui_2020, Helm_2024} or constructing synthetic benchmarks~\cite{Reif_2019}---without pinpointing, reproducing, or localizing the cause of each individual missed edge.

\item We improve the transparency of static analysis tools by surfacing unsoundness during execution, so users can judge the reliability of a generated call graph and respond.

\item We show that the same reports give coding agents a new form of guidance---beyond functional correctness---toward software that is more amenable to static analysis.

\end{itemize}

\vspace{-3pt}
\section{Motivating Example}
\label{sec:motivation}

Consider the program in Listing~\ref{lst:reflect_var}, which uses \texttt{Class.forName()} with a non-constant class name (line 4). When OPAL constructs the static call graph for this program with full reflection settings, it fails to resolve the reflective instantiation and consequently misses the edge from \texttt{ReflectVar.main()} to \texttt{Target.hello()}. We use this example throughout the paper, both to illustrate the problem and to walk through how \tool addresses it. A missed edge, however, has different consequences---and calls for different remedies---depending on whether one is a \emph{user} of OPAL or one of its \emph{developers}. We consider each in turn, developing goals that motivate the design of \tool.

\vspace{-3pt}
\subsection{User Perspective}

Suppose this program belongs to a security analyst who relies on OPAL for tasks such as vulnerability analysis or providing structural context to an LLM. The analyst receives no indication that the call graph is incomplete, and so may assume the omitted methods are unreachable and base critical decisions on an unsound result. Consulting documentation and prior studies yields only high-level statistics---e.g., Reif et al.~\cite{Reif_2019} report that OPAL resolves only 6 of 16 reflection test cases---which say nothing about whether the specific pattern in \emph{this} program is supported. Suppose instead the tool warned the user on encountering a potential source of unsoundness in their program: for Listing~\ref{lst:reflect_var} it would reveal a reflective call with a non-constant class name that OPAL cannot resolve, and the user could then switch tools, apply a complementary technique such as Tamiflex~\cite{tamiflex}, or refactor to avoid the pattern.

This motivates our first goal, \textbf{(G1) User transparency:} during the analysis of a user's own program, surface every pattern the tool cannot model soundly, with enough explanation for the user to assess the risk and respond.

\begin{fixedlisting}{Reflection with non-constant class name.}{lst:reflect_var}
\begin{lstlisting}[style=javaStyle]
public class ReflectVar {
  public static void main(String[] args) {
    String className = "Target"; //variable
    Class<?> cls = Class.forName(className);
    Object obj = cls.newInstance();
    ((Target) obj).hello();
  }
}
class Target {
  public void hello() {
    System.out.println("Hello");
  }
}
\end{lstlisting}
\vspace{-3pt}
\end{fixedlisting}

\begin{fixedlisting}{OPAL's limitation in handling non-constant reflection arguments. The unmodified lines are OPAL's own implementation; the added (\textcolor{diffadd}{green}) lines are the unsoundness warning \tool inserts (Section~\ref{sec:tool_annotation}).}{lst:tool_limitation}
\begin{lstlisting}[style=scalaDiffStyle]
 AllocationsUtil.handleAllocations(value, context, depender, stmts, _ eq ObjectType.String, failure) {
   (_, defSite, _stmts) =>
     getString(defSite, _stmts) match {
       case Some(v) => strings += v
       case _ =>
+        // ---- UNSOUNDNESS WARNING ----
+        System.out.println("[UNSOUNDNESS] | non_constant_reflection" + " | caller=" + context + " | site=" + depender)
         failure()
     }
 }
 strings

 private[reflection] def getString(stringDefSite: Int, stmts: Array[Stmt[V]]): Option[String] = {
   val expr = stmts(stringDefSite).expr
     expr match {
       case StringConst(_, v) => Some(v)
       case _                 => None
     }
 }
\end{lstlisting}
\vspace{-3pt}
\end{fixedlisting}

\vspace{-3pt}
\subsection{Tool Developer Perspective}

Now consider the same missed edge from an OPAL developer's perspective. Its source could be a deliberate design decision, an implementation bug, or an unrealized opportunity to improve soundness---but the tool offers no signal that tells these apart: it neither flags the omission nor indicates where, in its own implementation, the edge was dropped.

Suppose instead the developer were handed exactly the information needed to act on it. Existing work stops short: a report in the style of Sui et al.~\cite{Sui_2020} would attribute the missing edge to a reflective API on the execution path, such as \texttt{Class.newInstance()} (line 5)---symptom-level attribution that points to the right neighborhood but is neither precise enough to localize the limitation nor specific enough to reproduce it reliably.

Three further pieces of information would close this gap: the precise condition that triggers the unsoundness, the responsible code location, and a reproducing test case. Inspecting OPAL's implementation (Listing~\ref{lst:tool_limitation})\footnote{The excerpt is from OPAL v5.0.1, the version used throughout our evaluation.} reveals the condition: the true cause lies earlier in the execution chain than \texttt{Class.newInstance()}---the argument to \texttt{Class.forName()} is a non-constant variable. At line~17, OPAL pattern-matches the argument against \texttt{StringConst} and returns \texttt{None} for any non-constant argument, so the reflective target, and every edge depending on it, is never resolved. Reflection itself is not the problem; reflection \emph{with a non-constant class name} is. For the test case, Listing~\ref{lst:reflect_var} is exactly such a case, with the virtual call at line~6 the expected edge against which both the bug and any fix can be confirmed.

Without such support, a developer must hypothesize the cause, build a reproducing test case, and localize the code by hand---demanding substantial expertise in both the framework and the language feature. This motivates our second goal, \textbf{(G2) Developer localization:} for each missed edge, automatically explain the precise root cause, the responsible tool implementation location, and a reproducing test case with its expected edge, so that the developer can confirm, fix, or knowingly accept the limitation.

\vspace{-3pt}
\section{\tool Approach}
\label{sec:approach}



\begin{figure*}[t]
    \centering
    \includegraphics[width=\textwidth]{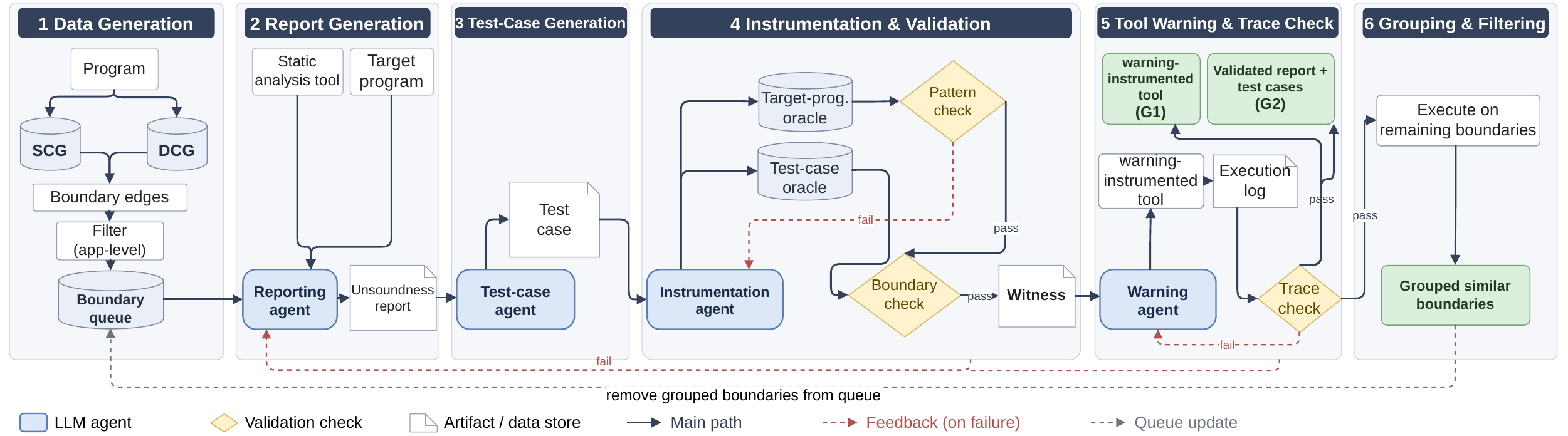}
    \caption{\Training phase of \tool. From each boundary edge---a call observed at runtime but absent from the static call graph---the reporting agent suggests a root cause, which \tool then checks by collecting three independent pieces of evidence: runtime instrumentation of the program, a reproducing test case, and a warning inserted into the analysis tool.}
    \label{fig:workflow}
\end{figure*}

This section presents \tool, an agentic framework for finding the root causes of unsoundness in call graph construction tools.
\tool's \training phase is given a \emph{benchmark} of target programs, and for each follows the workflow detailed in Figure~\ref{fig:workflow}, which is divided into six stages. The first stage, \emph{data generation} (Section~\ref{sec:data_generation}), collects \emph{boundary edges}---calls observed when the program runs but missing from its static call graph---by building the program's static and dynamic call graphs, taking their difference, and discarding non application-level edges. Next, for each boundary edge, the reporting agent (\emph{report generation}, Section~\ref{sec:report_generator}) inspects the program and the tool's source to suggest \emph{why} the edge was missed---which pattern in the program the analysis failed to model, and where in the tool's implementation that failure occurs---producing an unsoundness report with a root cause, the implicated tool location, and a test-case description. Next, \tool collects evidence to validate this report, in three independent forms. The test-case agent (\emph{test-case generation}, Section~\ref{sec:test_case_generation}) turns the report's description into a test case that reproduces the missing edge. The instrumentation agent (\emph{instrumentation and validation}, Section~\ref{sec:oracle_generator}) writes runtime instrumentation that detects the suggested pattern as the program executes; \tool checks that it flags the original boundary edge and that the test case reproduces the same pattern, producing a \emph{witness}. The warning agent (\emph{tool warning and trace check}, Section~\ref{sec:tool_annotation}) inserts a warning at the implicated code location in the tool that fires when the tool processes the pattern. When the three agree, \tool emits a validated report; when a check fails, it returns targeted feedback to the responsible agent and retries, setting the boundary aside after a bounded number of attempts. A final stage, \emph{grouping and filtering} (Section~\ref{sec:grouping}), reuses the instrumentation to dequeue and group other boundary edges sharing the same root cause.

These three pieces of evidence are the artifacts the two goals of Section~\ref{sec:motivation} call for. The test case, its expected edge, and the localized tool location serve the developer (\textbf{G2}), while in deployment the validated in-tool warnings fire on the same patterns in future programs to serve the user (\textbf{G1}).

\subsection{Data Generation}
\label{sec:data_generation}

For each target program we construct two call graphs: the static call graph, built by the target tool under the configuration we are studying, and a dynamic call graph, obtained by executing the program on representative inputs. We build both with the Judge framework~\cite{Reif_2019, Helm_2024}, which provides infrastructure for call-graph extraction and for comparing static and dynamic call graphs. The dynamic call graph is recorded by Judge's own execution machinery, which is unrelated to the instrumentation \tool later generates to validate a root cause (Section~\ref{sec:oracle_generator}).

Comparing the two graphs yields the \emph{boundary edges}: edges that appear in the dynamic call graph but not the static one, and whose caller appears in both.

Each boundary edge seeds the rest of the pipeline. Both goals G1 and G2 concern unsoundness at the \emph{application} level---patterns a user writes (G1) and a developer can reproduce as an application-level test case (G2)---so we keep only application-level boundary edges and exclude two other kinds. The first is edges whose caller is a JDK library method (e.g., in \texttt{java.*}): these lie within library infrastructure that users do not modify and a developer cannot readily isolate as a test. The second is edges whose callee is a \emph{JVM runtime} method---one the JVM invokes implicitly (e.g., \texttt{ClassLoader.loadClass()}, \texttt{<clinit>}, or the hidden classes generated for lambdas) rather than through a call site written in the program. Such edges are not stable, application-attributable patterns---which application method appears as the caller depends on the execution path---so localizing them is not meaningful. The complete list of excluded JVM runtime methods appears in our supplementary artifact~\cite{artifact}.

For the boundary edges we keep, we then collect the additional information the agents use to reason about them. First, the caller and callee methods with their full signatures, and the line number and program counter of the call site at which the edge was observed dynamically. Second, the execution context leading to that call site: from the program entry point we perform a depth-first search over the dynamic call graph to enumerate paths ending at the caller, recording each as its sequence of methods. Because many paths may reach the same caller and individual paths may be long, we keep at most $i$ paths ($i=5$) and truncate each to its last $j$ methods ($j=15$)---the methods nearest the call site, which are the most informative for reasoning about the cause. Third, when the call site is resolved statically but none matching the dynamic callee, we collect the targets the tool did resolve; this shows the agent how the tool interpreted the call site, which often points directly at why the expected callee was dropped.


\subsection{Unsoundness Report Generation}
\label{sec:report_generator}

Boundaries are processed one at a time from the queue. The first stage, \emph{report generation}, is carried out by the \emph{reporting agent}, an LLM agent whose task is to suggest the root cause of the missed edge. It has access to three inputs: (1) the source code of the static analysis tool, (2) the source and byte code of the target program for the current boundary, and (3) the boundary information from Section~\ref{sec:data_generation}.

The reporting agent starts from the boundary information. It uses the caller and callee signatures to locate the call site in the program and the recorded execution paths to gather the surrounding context, and from these it suggests the program \emph{pattern} responsible for the missed edge---the specific construct, and the condition under which the analysis fails to model it. In the motivating example, it identifies the pattern as a reflective call through \texttt{Class.forName()} whose class-name argument is non-constant (line 4); reflection alone is not the pattern, reflection with a non-constant argument is. The reporting agent then examines the tool's implementation to determine what it fails to handle for that pattern, refining its account of the limitation. For navigating both code bases it has access to the \texttt{grep}, \texttt{glob}, \texttt{read}, \texttt{write}, and \texttt{bash} tools.

\begin{lstlisting}[style=jsonStyle, float=t, caption={Example of unsoundness report.}, label={lst:opal_report}]
{"relevant_files": [
  {"file": "FastCharsetProvider.java",
  "lines": "108-119",
  "description": "Class name resolved via map lookup and used in Class.forName"
  }],
"tool_limitation": {
  "description": "String analysis only handles constant strings and cannot resolve class names derived from variables (e.g., map lookups) when analyzing reflective calls",
  "relevant_files": [
    "StringUtil.scala (71-77)",
    "TypesUtil.scala (29-46)",
    "ReflectionRelatedCallsAnalysis.scala (254-261)"
  ]},
  "potential_test_case": {
    "name": "ReflectVar",
    "description": "Main method invokes Target.hello where receiver is created via Class.forName with a non-constant (variable) argument."}
}
\end{lstlisting}

The output of the reporting agent is an unsoundness report for a specific boundary edge in JSON format. The report consists of three components: (1) source files in the target program that contribute to the identified root cause, along with brief descriptions; (2) the tool limitation responsible for the missed edge, including a one-sentence description and pointers to the relevant implementation files and code regions; and (3) a test case description designed to reproduce the issue. The unsoundness report in Listing~\ref{lst:opal_report} was generated after analyzing a boundary edge in program \texttt{SPM3} from the Cats benchmark, and its test case description corresponds to the motivating example in Listing~\ref{lst:reflect_var}.

The report is not necessarily final. If a later validation stage fails (Sections~\ref{sec:oracle_generator} to \ref{sec:tool_annotation}), the reporting agent is re-invoked with feedback describing the failure so it can revise its account of the root cause. To bound cost, it performs at most $k=2$ such revisions before \tool sets the boundary aside and moves to the next one in the queue.

\subsection{Test Case Generation}
\label{sec:test_case_generation}

From the unsoundness report, the \emph{test-case agent} constructs a small standalone program intended to reproduce the missed edge. It extracts the test-case description from the report, generates a corresponding program, and then builds and runs it, refining until it compiles and executes successfully. At this stage the goal is only a working, self-contained reproduction at the application level; whether the program actually exhibits the missed edge is established by the validation in Section~\ref{sec:oracle_generator}. If the agent cannot produce a compilable, runnable test case, \tool sets the boundary aside and moves to the next one in the queue.


\subsection{Instrumentation and Validation}
\label{sec:oracle_generator}

The reporting agent has only \emph{suggested} a root cause; \tool now gathers evidence for it. Two things must hold: that the suggested pattern accounts for the real missed edge (the boundary edge on the original target program), and that the generated test case reproduces that same pattern, making it a faithful, minimal stand-in. \tool establishes both by building \emph{instrumentation} that detects the pattern as the program runs.

The \emph{instrumentation agent} constructs this instrumentation from the report. To make the pattern precise enough to detect at runtime, we describe it with three elements, which the agent extracts from the report:

\paragraph{Root cause location} The program location where the unsoundness originates; the instrumentation must flag the corresponding instruction under the relevant runtime conditions. In the motivating example this is the \texttt{Class.forName()} call at line 4, where the argument is a variable rather than a constant.

\paragraph{Symptoms} The observable consequences of the root cause---for call graphs, the missed edges. In the motivating example the symptom is the virtual call at line 6 whose receiver originates from the reflective instantiation.

\paragraph{Relation} The link between root cause and symptom, typically a data-flow dependency. In the motivating example it is the flow from the object (\texttt{cls}) created by the reflective call at line 4 to the receiver of the virtual call at line 6.

Running the instrumented program produces a \emph{witness}: a JSON record listing each detected root cause together with the symptoms it produced. Each root-cause entry names the enclosing method and program counter of the root-cause instruction; each symptom entry names the caller, callee, and program counter of a missed edge. For the motivating example, the witness (Listing~\ref{lst:oracle_example}) records a single root cause---the \texttt{Class.forName()} call in \texttt{ReflectVar.main()}---and a single symptom, the missed edge from \texttt{ReflectVar.main()} to \texttt{Target.hello()}.

\begin{lstlisting}[style=jsonStyle, float=t, caption={Example witness produced by the instrumentation for the motivating example in Listing~\ref{lst:reflect_var}.}, label={lst:oracle_example}]
{"RC:ReflectVar.main(Ljava/lang/String;)V@2":
 {"method": "ReflectVar.main([Ljava/lang/String;)V",
  "pc": 1,
  "symptoms": [
   {"caller": "ReflectVar.main([Ljava/lang/String;)V",
    "callee": "Target.hello()V",
    "pc": 4}
  ]}
}
\end{lstlisting}

The instrumentation agent works iteratively, generating a candidate and then building and running it until it compiles and runs without error. If it cannot produce working instrumentation, \tool sets the boundary aside and moves on; otherwise, \tool applies the instrumentation in the two checks below.

\setcounter{paragraph}{0}
\paragraph{Check 1: the pattern explains the real missed edge} \tool runs the instrumentation on the \emph{original target program} and inspects the resulting witness. The check passes if the witness's symptoms include the boundary edge we started from---evidence that the suggested pattern is indeed what caused the tool to miss that edge.

\paragraph{Check 2: the test case reproduces the pattern} \tool runs the instrumentation on the \emph{test case} to obtain its witness. If that witness contains at least one root cause and symptom, the pattern is present in the test case. \tool then computes the boundary edges of the test case itself and confirms that the symptoms the witness predicts---restricted to those reachable in the static analysis---appear among them. Passing both parts shows that the test case contains the pattern and reproduces the missing edge, and it lets the witness single out the relevant missed edge among any others the test case might exhibit.







When the first check fails, \tool returns to the instrumentation agent for up to $k=3$ refinements, in case the fault lies in the instrumentation rather than the report. If the check still fails, \tool returns to report generation with feedback, including the instrumentation and the witnesses gathered so far, so that the reporting agent can revise the root cause. If the second check fails, \tool returns directly to the reporting agent to revise the root cause. Once both checks pass, the workflow proceeds to the next phase.

\subsection{Tool Warning and Trace Check}
\label{sec:tool_annotation}

The \emph{warning agent} inserts a \emph{warning} into the source of the static analysis tool at the implicated location---a print statement that, during call graph construction, emits a message whenever the tool reaches that location under the conditions of the root-cause pattern. Once inserted, the same warning fires on any future program that exercises the pattern, which is how \tool surfaces unsoundness to users (Section~\ref{sec:motivation}). Because the report from Section~\ref{sec:report_generator} already names the files and locations responsible, the agent edits only those.

Each warning records the description of the unsoundness, the enclosing method in the tool where the pattern is handled, and the line of that location. This information is usually available from variables in scope at the insertion point; when it is not, we instruct the agent to thread the needed variables into scope so the warning can report them.


The warning agent is given the unsoundness report and write access to the tool's source. The added (green) lines in Listing~\ref{lst:tool_limitation} show the warning for the motivating example, inserted directly into OPAL's implementation. When \texttt{handleAllocations} calls \texttt{getString}, the latter returns \texttt{None} for a non-constant argument, so the second case branch runs and triggers the inserted warning. During call graph construction, the message is emitted whenever the pattern is encountered.

With the warning in place, \tool runs the warning-instrumented tool on the test case and records its execution trace. The check passes if the trace contains the inserted warning and, further, the root-cause locations recorded in the test case's witness that are reachable in the static analysis all appear in the trace. This confirms that the implicated location is the one actually responsible for the missed edge.

The check can fail in two ways: the trace contains no warning at all, meaning execution never reaches that region of the tool; or it contains warnings but not the root causes the witness expects. In either case \tool re-invokes the warning agent up to $k=3$ times with feedback identifying the failure mode and the warnings already tried. As in the previous phase, if the check still fails, \tool returns to report generation, because a warning that cannot be made to fire at the implicated location is itself evidence that the report's localization is wrong; a boundary that survives neither the warning retries nor a report revision is set aside. When the check passes, \tool records the inserted location in the report and proceeds.

After every boundary in the queue has been processed, \tool merges all inserted warnings, using Git, into a single warning-instrumented build of the tool. This build can be run on fresh programs to alert users where their generated call graphs may be missing edges, realizing G1.

\subsection{Grouping and Filtering}
\label{sec:grouping}

Once a boundary edge has been fully validated, the instrumentation built for it (Section~\ref{sec:oracle_generator}) can recognize the same pattern cheaply in other benchmark programs, which is what keeps the pipeline affordable: rather than a full agent analysis of every remaining boundary, \tool reruns that instrumentation on the programs whose boundaries are still in the queue. This works because the boundary edges come from executions we can rerun, so the instrumentation observes directly which queued boundaries exhibit the pattern it was built to detect. If a queued boundary appears among its symptoms, it very likely shares the same root cause, so \tool groups it with the analyzed boundary and removes it from the queue. This is a heuristic---a shared symptom is strong evidence but not proof of an identical cause---so a grouped boundary can still be analyzed on its own if later evidence warrants.

\section{Evaluation}
\label{sec:evluation}
Our evaluation asks whether \tool meets the two needs set out in Section~\ref{sec:motivation}: user transparency (\textbf{G1}) and developer localization (\textbf{G2}). Both depend on the \training phase first producing artifacts that can be trusted, so we begin there and then take each persona in turn.

\textbf{RQ1 (\Training phase).} Does \tool produce correct, validated root-cause artifacts, and at what scale and cost? This establishes whether the engine that serves both personas is sound.

\textbf{RQ2 (Developers, G2).} Do \tool's reports, test cases, and localizations help tool developers diagnose, confirm, and fix unsoundness?

\textbf{RQ3 (Users, G1).} Do the warnings learned during \training generalize to previously unseen programs, giving a user a reliable signal of where their own call graph may be unsound?

\textbf{RQ4 (Users, G1).} Can a user act on \tool's reports---here, through a coding agent---to refactor a program so that it is more amenable to static analysis?

\subsection{Experimental Setup}
\label{sec:setup}

This subsection covers what is common to all four RQs; details specific to a single RQ appear in its subsection.

\paragraph{Tools} We apply \tool to three state-of-the-art call graph construction frameworks: OPAL~\cite{opal} (v5.0.1), WALA~\cite{wala2015} (v1.5.7), and Soot~\cite{soot} (v4.4.1), all used extensively in recent studies~\cite{Reif_2018, Helm_2024}.\footnote{To ensure that it is not overfit to a particular framework, we only experimented with WALA and OPAL as we developed \tool, and held Soot out to be only an evaluation target.} For each we use the default context-insensitive algorithm with all available soundness-enhancing options enabled. The three frameworks differ considerably in the size and structure of their call graph construction code---2666, 1997, 3804 files and roughly 300, 309, 705 KLoC for OPAL, WALA, and Soot---which bears on how hard it is for an agent to locate a root cause within each, a point we return to in RQ1.

\paragraph{Benchmarks} We use three complementary benchmarks.
\emph{Cats}~\cite{Reif_2019} is a suite of 112 small, manually written programs,
each exercising a specific dynamic language feature known to challenge call
graph construction; it lets us test \tool on isolated, well-understood patterns.
\emph{NJR}~\cite{njr_bench} and \emph{XCorpus}~\cite{xcorpus_bench} are
real-world applications that exercise the intertwined sources of unsoundness
arising in practice: for NJR we use the 141-program subset from prior
static-analysis studies~\cite{Utture_2022, Le-cong_2022}, and for XCorpus the
subset gathered by Sui et al.~\cite{Sui_2020}---the only programs with at least
one built-in test case, which we use to generate the dynamic call graphs.

A program enters our evaluation only if it has at least one boundary edge
surviving the filtering of Section~\ref{sec:data_generation}, since otherwise
\tool has nothing to analyze. The two kinds of benchmarks are then sampled
differently. Because each Cats program targets a distinct pattern, we evaluate
every tool on all of its own qualifying Cats programs---15 for WALA, 19 for
OPAL, and 21 for Soot---so each tool is exercised on every pattern it exposes.
For NJR and XCorpus we instead keep only the programs common to all three
tools---25 NJR and 8 XCorpus---so the tools are compared on an identical set.
From the 8 XCorpus programs we study three spanning a range of domains and
unsoundness sources: \texttt{javacc5.0}, a parser generator;
\texttt{fitjava-1.1}, an acceptance-testing framework; and
\texttt{nekohtml-1.9.14}, an HTML parser.

\paragraph{Platform and cost} We build the static and dynamic call graphs with the Judge framework~\cite{Reif_2019, Helm_2024} (Section~\ref{sec:data_generation}) on Java~8, allowing three hours for static and one hour for dynamic construction. \tool itself uses the Claude Code agentic framework with Opus 4.8 as the reasoning model, run under a Claude Code Max subscription (\$200 per month), which was sufficient to complete the entire evaluation. Analyzing one boundary edge took 30 minutes on average, varying with the size and complexity of the target program and tool.



\subsection{RQ1: Validity, Yield, and Cost of the \training Phase}
\label{sec:rq1}

RQ1 asks whether the \training phase can be trusted: whether the root causes it reports are \emph{valid}, how many of the boundary edges it manages to resolve---its \emph{yield}---and at what cost. We keep validity and yield separate because a high yield means little unless the resolved cases are also correct. 

\paragraph{Yield} Table~\ref{tab:rq1_results} (columns 3-10) reports, for each benchmark and tool, the total number of boundary edges and how \tool handled them. The total is itself a coarse measure of a tool's unsoundness: for a given program the dynamic call graph is the same across tools, so the boundary count reflects how much of it the tool's static graph missed---and, as the table shows, the tools differ here. In all, \tool encountered 1459 boundary edges. We call a boundary \emph{resolved} when either an agent reasoned about it directly and passed all three validation checks of Section~\ref{sec:approach} (151, 10.3\%), or the grouping stage (Section~\ref{sec:grouping}) matched it to an already-validated pattern (1227, 84.1\%). 
Grouping accounts for most resolved boundaries, which is what lets the \training phase scale past a thousand edges without a separate agent analysis of each. Grouping is highly concentrated: across all three tools the largest group alone covers 22–34\% of grouped edges and the top three cover almost half, while the rest tail off to a median group size of 2–4, so 25–35\% of reasoned boundaries account for 80\% of all grouping. 

Yield tracks program complexity. The small, single-feature Cats programs were resolved completely for all three tools, because their unsoundness is localized and an agent needs little surrounding context to reason about it. The real-world NJR and XCorpus programs were harder: there a root cause often spans several classes, packages, or third-party libraries, so the agent must reason over long-range dependencies, and both root-cause identification and artifact generation become more error-prone.

\paragraph{Failed cases} The pipeline is sequential---report, then test case, then instrumentation with the \emph{pattern} and \emph{boundary} checks, then warning with the \emph{trace} check---so a boundary that fails an early stage never reaches the later ones, and we categorize each failure by the last stage it reached (columns 6--9 of Table~\ref{tab:rq1_results}). Most failures were pattern- or boundary-check failures: the generated instrumentation did not explain the real missed edge, or the generated test case did not reproduce it. Both concentrate on the larger programs and more elaborate patterns, where two known weaknesses of LLM agents compound---writing instrumentation that runs correctly on real-world code, and reasoning over long-range dependencies. A few cases failed earlier still, when an agent could not produce a required artifact at all.

\begin{table*}[t]
\centering
\caption{Summary of the results produced by \tool for each benchmark and static analysis tool, and breakdown of validation feedbacks used in resolved cases. Failed cases are categorized by the last validation step in which they failed. * indicates a subset of the benchmark.}
\label{tab:rq1_results}
\begin{tabular}{ll|rrrrrrrr|rrrrr}
\toprule
\multirow{2}{*}{\textbf{Programs}} &
\multirow{2}{*}{\textbf{Tool}} &
\multirow{2}{*}{\textbf{Total}} &
\multicolumn{2}{c}{\textbf{Resolved}} &
\multicolumn{5}{c}{\textbf{Failed}} &
\multicolumn{4}{c}{\textbf{Feedback}} \\
\cmidrule(lr){4-5}
\cmidrule(lr){6-10}
\cmidrule(lr){11-14}
&&&
\textbf{Reasoned} &
\textbf{Grouped} &
\textbf{Agent} &
\textbf{Pattern} &
\textbf{Boundary} &
\textbf{Trace} &
\textbf{Total} &
\textbf{Pattern} &
\textbf{Boundary} &
\textbf{Trace} &
\textbf{Total}\\
\midrule

\multirow{3}{*}{Cats}
& WALA  & 18 & 15 & 3 & 0 & 0 & 0 & 0 & 0 & 0 & 0 & 8 & 8 \\
& OPAL  & 34 & 12 & 22 & 0 & 0 & 0 & 0 & 0 & 0 & 0 & 5 & 5 \\
& Soot  & 53 & 19 & 34 & 0 & 0 & 0 & 0 & 0 & 3 & 0 & 17 & 20 \\
\midrule

\multirow{3}{*}{NJR*}
& WALA  & 223 & 22 & 177 & 2 & 14 & 8 & 0 & 24 & 3 & 2 & 9 & 14 \\
& OPAL  & 448 & 29 & 386 & 1 & 18 & 14 & 0 & 33 & 2 & 5 & 9 & 16 \\
& Soot  & 548 & 46 & 485 & 0 & 10 & 5 & 2 & 17 & 7 & 2 & 41 & 50 \\
\midrule

\multirow{3}{*}{XCorpus*}
& WALA  & 18 & 1 & 16 & 0 & 1 & 0 & 0 & 1 & 0 & 0 & 0 & 0 \\
& OPAL  & 59 & 3 & 55 & 0 & 1 & 0 & 0 & 1 & 1 & 0 & 0 & 1 \\
& Soot  & 58 & 4 & 49 & 0 & 3 & 2 & 0 & 5 & 0 & 0 & 6 & 6 \\
\midrule

\multicolumn{2}{l}{\textbf{Overall}}
& \textbf{1459}
& \textbf{151}
& \textbf{1227}
& \textbf{3}
& \textbf{47}
& \textbf{29}
& \textbf{2}
& \textbf{81}
& \textbf{16}
& \textbf{9}
& \textbf{95}
& \textbf{120} \\

\bottomrule
\end{tabular}
\end{table*}







\paragraph{Validity} A boundary passing all three checks shows only that its report, test case, instrumentation, and warning are mutually consistent---not that the reported root cause is the true one. To test correctness directly, we manually audited the 151 reasoned boundaries---those an agent analyzed directly---confirming for each whether the reported pattern is the actual reason the edge was missed and whether the inserted warning fires at the responsible location. The root cause was correct in 100\% of audited cases.

However, we found two problems:
(1) The generated instrumentation is sometimes overly specific to the program
on which it was built and does not fire on others. Grouping is therefore
occasionally imperfect: a single root-cause pattern can be split across
boundaries and reasoned about more than once. Only 30/44, 39/69, and 30/38
reasoned boundaries correspond to unique root causes for OPAL, Soot, and WALA,
respectively. Because grouping is only a scalability heuristic
(Section~\ref{sec:grouping}), this does not affect resolved cases; it merely
makes the grouped count underestimate reuse. These counts also approximate the
number of fixes needed to address the identified unsoundness patterns.
(2) Some warnings are inserted at locations that are too general. For example, warnings associated with unmodeled native methods may be triggered whenever a receiver has an empty points-to set, even when the reported root cause is unrelated to the actual unsoundness. While these warnings still alert users to potential incompleteness, the explanation may be less actionable when attempting to eliminate the underlying root cause (Section~\ref{sec:rq4}). 
To mitigate this issue, we instruct the agent to guard warnings with lightweight control-flow conditions when possible, reducing false alarms from frequently executed but unrelated code locations.
Neither problem affects the 100\% correctness above: each reported root cause is still a genuine limitation of the tool that explains its missed edge; the problems only limit how broadly the instrumentation and warnings apply.

\paragraph{Grouping accuracy} In addition to the problems above, we validated whether grouping is too general: whether grouped edges actually share the root cause of the validated boundary that grouped them (the group's \emph{representative}). We used a stratified random sample of 100 of the 1227 grouped edges. We stratified first by tool and then by group: every group with at least 12 members forms its own stratum, the remaining small groups of each tool are pooled into one stratum, and samples are allocated in proportion to stratum size (WALA 20, OPAL 36, Soot 44). Each sampled edge was then re-analyzed from scratch by \tool with grouping disabled, and we compared the root cause it confirmed with the representative's. Overall, 86 of the 100 sampled edges share their representative's root cause (WALA 17/20, OPAL 31/36, Soot 38/44). The 14 mismatches are not spread evenly: they fall in 6 of the 109 total groups, mostly mid-sized and large ones, where the representative’s instrumentation was too general and accepted edges outside its root cause. Nevertheless, each of the 14 mismatched edges belongs to another existing group, so no new root causes were missed in this sample. With 86\% of sampled grouped edges sharing their representative’s root cause, and no new root causes missed among the mismatches, grouping is a valid strategy that let \tool scale to over a thousand boundaries from only 151 direct agent analyses.

\paragraph{Feedback loops} The last four columns in Table~\ref{tab:rq1_results} show, for the resolved boundaries, how often each check first failed and was then repaired by the feedback returned to the responsible agent---a measure of how much the iterative loops contribute. Out of 151 reasoned boundaries, 83 of them were resolved by at least one round of feedback (55\%) with an average of 1.8 rounds for each boundary. This is consistent with two factors: Soot had the most reasoned boundaries, and its call graph construction code is larger and more intricate than that of the other tools (Section~\ref{sec:setup}). Some of this feedback goes back to the unsoundness report itself. When one piece of evidence (the instrumentation, the test case, or the warning) held but another could not be satisfied, the framework returned the feedback to the report generator, which revised the root cause, the test-case description, or both. Of the 151 reasoned boundaries, 29 (19\%) needed such a restart. For example, in one WALA case, a missing call edge to \texttt{ArrayList.iterator()} arose because the receiver’s points-to set was empty. The initial report attributed this to reflection: the plugin object holding the list is created by \texttt{newInstance()} on a non-constant \texttt{Class}, which WALA does not model. The pattern check, however, failed to confirm this explanation, and its feedback sent the report back for revision. The revised report traced the empty set to an earlier point in the data flow: the plugin’s metadata is stored in a \texttt{ConcurrentHashMap} with \texttt{put} and later retrieved with \texttt{get}, but the 0-CFA builder installs no model for maps, so the stored value never reaches the retrieval. This explanation passed all checks and was confirmed.

\subsection{RQ2: Tool Developer Support}
\label{sec:rq2}

RQ2 asks whether \tool's artifacts---the report, the reproducing test case, and the implicated tool location---give a developer enough to act on an unsoundness. We assess this by sharing them with the maintainers of the three analyzed tools. 
Through manual inspection, we identified 7, 9, and 9 distinct root-cause groups for WALA, OPAL, and Soot, respectively. From these groups, we selected five from WALA, two from OPAL, and five from Soot and sent our artifact to the respective maintainers. 
For each report we record whether the maintainer confirmed it and, if so, whether they treated it as a bug to fix or as an intentional tradeoff.

Every confirmed limitation was judged actionable, and several led to concrete outcomes. Soot maintainers were the most responsive. They confirmed four of our five reports and fixed three of them: missing handling of implicit exceptions, dynamic proxies (\texttt{Proxy.newProxyInstance}), and reflective field access (\texttt{Class.getDeclaredFields}). The fourth, reflective calls with non-constant parameters, was scheduled to be addressed in a commercial version of the tool, and the fifth had not been discussed yet. 

The WALA maintainers confirmed two---reflective field access (e.g., \texttt{Class.getDeclaredField}) and reflective proxy methods (e.g., \texttt{Proxy.newProxyInstance})---and classified a third, implicit exceptions, as an intentional design tradeoff that they noted should be documented more clearly. That last case is evidence our artifacts are sufficient to distinguish a bug from a deliberate tradeoff, part of Goal G2. We did not receive a response for the other two reports. 

The OPAL maintainers confirmed one limitation, missing \texttt{System.out} edges when the JRE is in scope, which they had already fixed in a newer release. They did not respond to our remaining report. 

Even in cases developers did not respond, our artifacts let us carry out that diagnosis ourselves. The reports let us reason about a limitation the way a maintainer would, without their involvement. For example, as shown in Listing~\ref{lst:null_pointer_bug}, OPAL's own code notes that implicit \texttt{NullPointerException} calls should not be ignored, yet our report for a boundary involving such a call shows the limitation is unresolved---and that the location OPAL would handle it at is never reached for the affected cases. Because of imprecision in OPAL's handling of interface-typed receivers, the condition that would introduce the exception edge does not fire, so the expected edge to \texttt{NullPointerException.<init>} is never added. This points to an implementation bug rather than an intentional tradeoff.

\begin{fixedlisting}{Excerpt of the OPAL impl. of null receivers.}{lst:null_pointer_bug}
\begin{lstlisting}[style=scalaStyle]
call.receiver match {
  ...
  case _: IsNullValue =>
    // TODO: do not ignore implicit
    // NullPointerException.<init> calls
}
\end{lstlisting}
\end{fixedlisting}

Taken together---confirmed limitations, four fixes committed, a tradeoff the maintainers agreed to document, and a bug we localized unaided---these outcomes answer RQ2 in the affirmative: \tool's artifacts are actionable for developers. We intentionally selected a small, diverse set of reports to evaluate the actionability of \tool's artifacts rather than the prevalence of unsoundness issues. As a result, our evidence is qualitative and limited to the examined reports and the maintainers who chose to engage (Section~\ref{sec:threats}).

\subsection{RQ3: Generalizability to User Programs}
\label{sec:rq3}

We run three experiments to answer RQ3. The first two measure whether warnings learned from some programs fire on the unsoundness of other, unseen programs; the third asks whether microbenchmarks alone suffice to build those warnings.

\paragraph{First experiment: coverage of held-out programs}
We pool the programs from all three benchmarks and split them randomly into training and testing sets in an 80--20 ratio: the training programs drive the \training phase that produces the warnings, and the testing programs stand in for the deployment phase---future applications run through the warning-instrumented tool. To reduce the effect of how the split falls, we use five-fold cross-validation and report the average coverage and standard deviation across folds.

Re-running the warning-instrumented tool for every fold would be costly, so we use the equivalence groups from Section~\ref{sec:grouping} as a proxy. Recall that each group bundles boundary edges that the analysis confirmed share a single root-cause pattern. We say a group's warning is available if at least one of its edges originates in a training program---i.e., analyzing that program would have produced the warning---and we count a testing-program edge as covered if it belongs to an available group, since the same warning applies to every edge in the group. Coverage is the fraction of testing-program boundary edges that are covered.

Under five-fold cross-validation, the warnings cover 74.0\% $\pm$ 30.0 (WALA), 82.2\% $\pm$ 10.2 (Soot), and 92.3\% $\pm$ 5.1 (OPAL) of the boundary edges in held-out programs. This is the fraction of an unseen program's unsoundness that the warning-instrumented build would surface to its user with no dynamic analysis on the user's part (\textbf{G1}): a sufficiently diverse training subset captures most common causes and flags them even in programs that contributed nothing to the warnings.

\paragraph{Second experiment: running the warning-instrumented tools}
To validate the proxy and evaluate the instrumented tools directly, we selected eight NJR programs that were not part of our evaluation benchmark and that contain boundaries for all three tools after initial filtering. We first ran \tool on each program to establish ground truth, finding 6 root causes for WALA, 12 for Soot, and 11 for OPAL. We then ran the final warning-instrumented tools on the same programs and compared the emitted warnings against these root causes. The warnings cover 5 of 6 (83\%) root causes for WALA, 9 of 12 (75\%) for Soot, and 9 of 11 (82\%) for OPAL. These results are in line with the cross-validation estimates and confirm that the instrumented tools surface most of the unsoundness in programs they have never seen (\textbf{G1}).

\paragraph{Third experiment: microbenchmarks are not enough}
This experiment again uses the grouping proxy. Taking the root causes discovered in Cats as the only training set and the real-world NJR and XCorpus programs as the testing set, only 2.9\% of real-world boundaries match a Cats pattern. The rest come from patterns absent in the microbenchmark---typically interactions among application code, libraries, and frameworks---so a deployment built only from Cats would miss almost all of it. Diverse, real-world programs are thus essential to building the warnings future users rely on. We discuss remaining threats of the proxy methodology in Section~\ref{sec:threats}.

\subsection{RQ4: Generate Amenable to Static Analysis Code}
\label{sec:rq4}


Unsoundness can be removed by improving the analysis tool, but it can also be reduced by changing the program to avoid patterns the analysis cannot model soundly.\footnote{This is in effect what a sound static type system does---it restricts programs to a syntactic subset that can be analyzed soundly.} To investigate this possibility, we considered all 47 reasoned boundaries in NJR benchmark whose caller belongs to the application. For each, we give a coding agent the corresponding \tool report and instruct it to eliminate the root-cause pattern while preserving the program’s functionality. To validate each change, a separate agent generates tests covering the modified and affected regions, following established refactoring-validation practice~\cite{refactoring-validation}. A boundary is considered resolved when the refactored program passes these tests and the boundary edge no longer appears.
Overall, 30 of the 47 boundaries (64\%) could be resolved this way. The remaining 17 (36\%) fall into two categories: those involving APIs that cannot be safely replaced, and therefore require modeling in the analysis tool, and those whose behavior cannot be resolved statically and requires hybrid analysis. We present three successful cases.

\paragraph{Soot -- url088e9}

In this case, Soot fails to resolve call graph edges for \texttt{conn} due to imprecision in handling reflective class loading (Listing~\ref{lst:sootcase}). The analysis cannot infer the concrete type because \texttt{URL.openConnection()} ultimately relies on \texttt{Class.forName()} with a non-constant string, preventing Soot from adding the resulting HTTP connection implementation to the points-to set. Consequently, virtual calls such as \texttt{conn.setConnectTimeout()} remain unresolved.
Using the generated report, the coding agent replaces the reflective allocation path with an explicit allocation of the concrete connection type, making the allocation site visible to the analysis and enabling correct call resolution.
This single modification recovers 31 previously missed edges in the application that may be critical in networking applications. 

\begin{lstlisting}[style=diffStyle, float=tb, caption={Soot -- url088e9 case study.}, label={lst:sootcase}]
 URL url = new URL(BaseURL.getUrlForPath(path, enableSSL));
-conn = (HttpURLConnection) url.openConnection();
+conn = new sun.net.www.protocol.http.HttpURLConnection(url, Proxy.NO_PROXY);
 conn.setConnectTimeout(DEFAULT_CONNECTION_TIMEOUT);
\end{lstlisting}




\paragraph{WALA -- url3426a}

In this case, the application initializer reads the protected static field \texttt{AbstractDNSClient.DEFAULT\_CACHE} through reflection, by calling \texttt{getDeclaredField}, \texttt{setAccessible}, and \texttt{get} on the resulting \texttt{Field} object (Listing~\ref{lst:walacase}). The generated report identifies the root cause as the absence of a model for the native method \texttt{Class.getDeclaredFields0(boolean)} in WALA's \texttt{natives.xml}. As a result, WALA never sees the allocation of the \texttt{Field} object, so the calls made on it are left unresolved. Using this information, the coding agent replaced the reflective read with a plain virtual call to \texttt{getCache()} on the \texttt{DNSClient} instance allocated in the same initializer, which WALA resolves through ordinary call and field edges. This change resolved 4 missing edges.

\begin{lstlisting}[style=diffStyle, float=tb, caption={WALA -- url3426a case study.}, label={lst:walacase}]
 static {
   LRUCache cache = null;
   try {
-    Field f = AbstractDNSClient.class.getDeclaredField("DEFAULT_CACHE");
-    f.setAccessible(true);
-    cache = (LRUCache) f.get(null);
+    cache = (LRUCache) DNSCLIENT.getCache();
   } catch (...) {
     throw new IllegalStateException(e);
   }
   DEFAULT_CACHE = cache;
\end{lstlisting}

\paragraph{OPAL -- url1d5e}

In this case, OPAL does not model \texttt{EnumSet.allOf()}, preventing it from inferring the contents of \texttt{typeSet} (Listing \ref{lst:opalcase}). As a result, the points-to set of \texttt{typeSet} remains empty, causing OPAL to miss all virtual call edges involving the loop variable \texttt{type}. The generated report identifies this root cause and suggests replacing the reflective collection construction with an explicit allocation visible to the analysis. After this modification, the analysis recovered 4 missing edges.

\begin{lstlisting}[style=diffStyle, float=tb, caption={OPAL -- url1d5e case study.}, label={lst:opalcase}]
-EnumSet<LegacyOCPMessageTypes> typeSet =
-  EnumSet.allOf(LegacyOCPMessageTypes.class);
+LegacyOCPMessageTypes[] typeSet = {
+  LegacyOCPMessageTypes.HEARTBEAT,
+  LegacyOCPMessageTypes.ARE_YOU_MASTER,...};
 for (LegacyOCPMessageTypes type : typeSet)
   if (type.getSupportRx()){...}
\end{lstlisting}

Together, the three cases above recover 39 missed edges, from 4 (WALA and OPAL) to 31 (Soot).
These results answer RQ4: a user can use the reports not only to gauge soundness but to improve it, refactoring a program into a form the analysis handles, although the gain depends on the pattern and the program.

\section{Threats to Validity}
\label{sec:threats}


\paragraph{Construct validity} Our ground truth for unsoundness is the gap between a static and a dynamic call graph, and the dynamic graph is only a lower bound on real behavior: it contains the edges some execution exercised, not every feasible edge. A boundary edge is therefore genuine unsoundness, but the absence of one does not certify soundness, and richer inputs would surface more. 
For the same reason, we do not report the precision of the warnings. A warning raised where no boundary edge was observed may still flag a real missed edge that no execution happened to exercise, so counting it as a false alarm would be unreliable; prior work shows that treating unobserved edges as infeasible leads to substantially wrong labels~\cite{Rafieian_2026}.
Relatedly, \tool attributes each missed edge to a single root-cause pattern; one can in principle arise from several interacting limitations, so a warning may over- or under-approximate the affected locations when causes combine. A single cause accounted for the majority of those we found, and multi-cause patterns would require the agent to add complicated logic to the tool---both supporting this design choice.

\paragraph{Internal validity} The pipeline relies on LLM agents, whose behavior is non-deterministic: a different run may reason differently about the same boundary. The three validation checks and the feedback loops guard against this---an unsupported suggestion does not pass---but they cannot rule out a suggestion consistent with all the evidence yet still wrong, which is why we audited all reasoned boundaries by hand (Section~\ref{sec:rq1}). That audit, and our classification of failures, rest on author judgment; to mitigate this, one author classified and the LLM confirmed the classification by reading the reports. Finally, RQ3 uses grouping as a proxy for running the warning-instrumented tool, and since grouping is an imperfect heuristic (Section~\ref{sec:rq1}) the reported coverage could be off in either direction. \tool also relies on LLM-generated instrumentation as a witness for the root cause (Section~\ref{sec:oracle_generator}): although we validated it manually and added multiple validation checks, it may still miss the actual pattern or produce evidence only indirectly related to the root cause.

\paragraph{External validity} Our findings come from three tools and three benchmarks and may not carry to others. We chose widely used frameworks and a mix of microbenchmark and real-world programs, and held Soot out during development to check the approach is not tuned to the tools we built it on; even so, other tools, languages, and program styles could exercise patterns we did not see. The RQ2 evidence is further limited to a small set of reports and the maintainers who engaged, and should be read as qualitative.

\section{Related Work}
\label{sec:related}

\paragraph{Imprecision and unsoundness root causes in CG analysis}

Over the past decade, several studies have investigated techniques for identifying the root causes of imprecision and unsoundness in static analyses and quantifying their impact.
Sui et al.~\cite{Sui_2020} proposed an approach for identifying the language constructs responsible for missed call graph edges in Java programs. Their technique combines calling-context trees~\cite{calling_context} with runtime tagging of reflective operations to determine which language features contribute to unsoundness. Chakraborty et al.~\cite{chakraborty_2022} extended this idea to JavaScript, where functions are first-class values, by tracing function data flow to identify the origins of missed call graph edges. 
Andreasen et al.~\cite{andreasen_2017} introduced techniques for isolating soundness and precision issues in the TAJS JavaScript analyzer. Their approach records runtime values during program execution and validates that the corresponding static abstractions account for those values. When unsoundness is detected, delta debugging~\cite{delta_debugging} is used to minimize the program, making manual root-cause analysis significantly easier.

Several studies have focused on understanding analysis imprecision rather than soundness. Lhoták~\cite{Lhotak_2007} compared static and dynamic call graphs to identify sources of imprecision in Java call graph construction. Wei et al.~\cite{wei_2016} leveraged dynamic execution traces to localize the causes of imprecision in JavaScript analyses and suggested possible fixes. Similarly, Lee et al.~\cite{lee_2020} constructed tracing graphs by propagating information backwards from imprecise program points, enabling users to identify the dominant causes of analysis imprecision. However, because dynamic call graphs under-approximate program behavior, using them to identify imprecision is challenging~\cite{Rafieian_2026}. 

Unlike imprecision, unsoundness can be identified reliably from runtime information: an edge observed at runtime but missing from the static call graph is a definite miss. Beyond this, our work differs in two key respects. First, existing approaches generally require manual inspection, labeling, or interpretation of their results, which limits scalability and the diversity of patterns identified; our framework performs the entire workflow automatically, from identifying boundary edges to generating reports, reproducing test cases, and localizing the implementation limitation. Second, \tool identifies root causes at a much finer granularity. Existing techniques typically attribute a missed edge to a broad language feature or API (e.g., reflection or \texttt{Method.invoke()}), but modern tools support subsets of these features, so not every use leads to unsoundness. Our approach pinpoints the specific pattern---for example, that the unsoundness arises only when the argument to \texttt{Class.forName()} is not a compile-time constant---yielding more precise explanations and artifacts directly actionable for both developers and users.

\paragraph{Empirical evaluation of call graph analysis}
A number of studies have evaluated the soundness and precision of static call graphs~\cite{Jens_2017, Helm_2024, Reif_2018, Reif_2019, Li_2018}. Reif et al.~\cite{Reif_2018, Reif_2019} introduced a framework for benchmarking static call graph construction and developed a synthetic benchmark suite consisting of small programs with known sources of soundness and precision issues that are commonly missed by existing static analyses. Since these benchmarks come with known ground truth, they enable precise evaluation of static call graph quality. Helm et al.~\cite{Helm_2024} extended this line of work by applying fuzzing to real-world programs to approximate ground-truth dynamic call graphs. By comparing these dynamic call graphs with statically generated ones, they evaluated the soundness of several static analysis tools on real-world applications. Samhi et al.~\cite{Samhi_2024} applied a similar methodology to Android applications, highlighting the challenges posed by reflection and other dynamic language features.

Our work also leverages the static--dynamic comparison, using it to identify the boundary edges that start our analysis. We then go beyond measuring or categorizing unsoundness: we automatically reason about the root cause of each missed edge, generate a reproducing test case, localize the implementation limitation, and add a warning that surfaces the pattern in future analyses---providing actionable artifacts for developers and users rather than merely reporting that unsoundness exists.


\section{Conclusions}

Starting from a set of static and dynamic call graphs produced for a benchmark of target programs, \tool isolates individual missed static edges and suggests a root cause for each with an LLM agent, accepting that suggestion only when runtime instrumentation, a reproducing test case, and an in-tool warning each agree. Its output serves two stakeholders: developers, who get a reproducible, localized account of each limitation; and users, who get a warning-instrumented build that flags the same patterns in their own programs, with no dynamic analysis required.
Our evaluation on WALA, OPAL, and Soot shows that this design delivers on both fronts: the root causes \tool reasons about hold up under manual audit and extend through grouping to most remaining boundaries; maintainers confirmed, and in several cases fixed, the limitations we reported; the warnings generalize to programs they were not learned from; and the reports guided a coding agent to make programs more amenable to static analysis.

While we study call graph construction, the approach is not specific to it. In future work, we plan to apply it to other static analyses, such as security-oriented tools like CodeQL~\cite{codeql2021}, to bring the same transparency to the findings their users rely on.



\paragraph*{AI Usage} This paper was authored in collaboration with Claude Opus, which provided structural advice and made grammatical and writing improvements.

\section*{Data Availability}
We have made our framework and results available~\cite{artifact}.

\section*{Acknowledgment}
This work was supported in part by the National Science Foundation under grant CCF-2047682.

\bibliographystyle{IEEEtran}
\bibliography{references}

\end{document}